\documentstyle[aasms4]{article}

\begin{document}
\lefthead{Huang et al.}
\righthead{New Wolf-Rayet Galaxies}

\def\ergscm{\ifmmode {\rm ergs\,s}^{-1}\,{\rm cm}^{-2}
	  \else ergs\,s$^{-1}$\,cm$^{-2}$\fi}
\def\ergscmA{\ifmmode {\rm ergs\,s}^{-1}\,{\rm cm}^{-2}\,{\rm \AA}^{-1}
	  \else ergs\,s$^{-1}$\,cm$^{-2}$\,\AA$^{-1}$\fi}
\newcommand{\totder}[2]{\frac{d{#1}}{d{#2}}}
\newcommand{\parder}[2]{\frac{\partial{#1}}{\partial{#2}}}

\newcommand{\gsim}{\mbox{\hspace{.2em}\raisebox{.5ex}{$>$}\hspace{-.8em}\raisebox{-.5ex}{$\sim$}\hspace{.2em}}}
\newcommand{\lsim}{\mbox{\hspace{.2em}\raisebox{.5ex}{$<$}\hspace{-.8em}\raisebox{-.5ex}{$\sim$}\hspace{.2em}}}
\newcommand{\beq}{\begin{equation}}     \newcommand{\eeq}{\end{equation}}
\newcommand{\bey}{\begin{eqnarray}}     \newcommand{\eey}{\end{eqnarray}}
\newcommand{\etal}{{\em et al.\/}}
\newcommand{\ie}{{\em i.e.\/}}  \newcommand{\eg}{{\em e.g.\/}}
\newcommand{\lt}{\left} \newcommand{\rt}{\right}
\newcommand{\ssst}{\scriptscriptstyle}
\newcommand{\E}[1]{\times 10^{#1}}

\def\arcsecpoint{\ifmmode ''\!. \else $''\!.$\fi}
\newcommand{\s}{\,{\rm s}}      \newcommand{\ps}{\,{\rm s}^{-1}}
\newcommand{\yr}{\,{\rm yr}}    \newcommand{\Hz}{\,{\rm Hz}}
\newcommand{\cm}{\,{\rm cm}}    \newcommand{\km}{\,{\rm km}}
\newcommand{\kmps}{\km\s^{-1}}
\newcommand{\parsec}{\,{\rm pc}}\newcommand{\Mpc}{\,{\rm Mpc}}
\newcommand{\um}{\,\mu{\rm m}}
\newcommand{\ergs}{\,{\rm ergs}}        \newcommand{\K}{\,{\rm K}}
\newcommand{\eV}{\,{\rm eV}}    
\def\Hubble{\ifmmode {\rm km\,s}^{-1}\,{\rm Mpc}^{-1} 
	\else km\,s$^{-1}$\,Mpc$^{-1}$\fi}
\newcommand{\za}{z_{1}}         \newcommand{\zb}{z_{2}}
\newcommand{\zo}{z_{0}}         \newcommand{\tha}{\theta_{app}}
\newcommand{\va}{v_{1}}         \newcommand{\vb}{v_{2}}
\newcommand{\Ys}{Y_{\ast}}
\newcommand{\cs}{\xi_{\ast}}    \newcommand{\ts}{\theta_{\ast}}
\newcommand{\Dy}{\Delta_{Y}}    \newcommand{\Dv}{\Delta_{v}}
\newcommand{\ro}{r_{\ssst 0}}
\newcommand{\HeII}{He {\sc ii} $\lambda 4686$}
\newcommand{\CIV}{C {\sc iv} $\lambda 5808$}
\newcommand{\NIII}{N {\sc iii} $\lambda 4640$}
\newcommand{\NV}{N {\sc v} $\lambda 4606$}
\newcommand{\Ha}{\mbox{H$\alpha$}}
\newcommand{\Hb}{\mbox{H$\beta$}}
\newcommand{\Hr}{H$\gamma$}
\newcommand{\HE}{H$\epsilon$}
\newcommand{\Hd}{H$\delta$}
\newcommand{\bc}{\begin{center}}
\newcommand{\ec}{\end{center}}

\title{New Wolf-Rayet Galaxies with Detection of WC Stars}

\author{J.H. Huang\altaffilmark{1,2}, Q.S. Gu\altaffilmark{1,2}, L. Ji\altaffilmark{1,2}, W.D. Li\altaffilmark{3}, J.Y. Wei\altaffilmark{3}
and W. Zheng\altaffilmark{4}}
\altaffiltext{1}{Department of Astronomy, Nanjing University, Nanjing, 210093, China}
\altaffiltext{2}{United Lab for Optical Astronomy, The Chinese Academy of Sciences, 100080, China}
\altaffiltext{3}{Beijing Astronomical Observatory, The Chinese Academy of Sciences, 100080, China}
\altaffiltext{4}{Department of Physics and Astronomy, The Johns Hopkins University, Baltimore, MD 21218-2686}

\begin{abstract}
We report the discovery of two new Wolf-Rayet (WR) galaxies: Mrk~1039, 
and F08208$+$2816. 
Two broad WR bumps at 5808\AA~ and 4650\AA~ indicate the presence of WCE and 
WNL star population in all two sources.
We also confirm the presence of WR features in Mrk~35, previously detected
in a different position.
The observed equivalent width of the WR bump at 4650\AA~
and the derived number ratios of WR/(WR$+$O) imply that star formation in 
these sources takes place in short burst duration. 
Comparisons with the recent models of WR populations in young starbursts
with the observed EW(\HeII)/EW(\CIV)/EW(WRbump) and their relative intensities
provide an indication that the stellar initial mass function in some WR
galaxies might not be Salpeter-like. 
It is interesting to find that the luminous IRAS source, F08208$+$2816, has 
little dust reddening, probably because of the existence of a 
powerful superwind. By comparisons with other starbursts observed with the 
Hopkins
Ultraviolet Telescope, F08208$+$2816 as a merging system renders a chance
to study the contribution from young starbursts to the UV background radiation
in universe.
\end{abstract}


\section{INTRODUCTION}

Wolf-Rayet (WR) galaxies are a subset of H~{\sc ii} galaxies with broad 
emission lines that are characteristic of WR stars (Conti 1991). Outside the local
universe, the broad feature usually detected in WR galaxies is the so-called
WR bump, a blend of He~{\sc ii} and other lines from C {\sc iii}, 
N~{\sc iii}/N~{\sc v}. Among about
90 known WR galaxies (Schaerer 1997, private communication),
thirteen show the broad \CIV\ line, a strong
feature from WC stars. Indeed, from the current knowledge of stellar
evolution one would expect to detect WC stellar population in about 1/3 of
all WR galaxies with a metallicity of (1/5 - 1)Z$_{\odot}$ (Schaerer \&
Vacca 1998, SV98 hereafter). The deficiency of such detection
might be caused by the weakness of \CIV\ emission, as compared with the WR
bump. 
Alternatively, uncertainties in our knowledge of the evolution of massive stars 
(e.g. their mass-loss rates, Meynet 1995) could be responsible for 
such an apparent deficiency.
For example, the recent discovery of WR stars in I Zw 18 (Izotov et al. 
1997$a$,
Legrand et al. 1997) seems to indicate enhanced mass-loss rates.
These uncertainties also complicate derivations of the upper end of the 
initial mass function (IMF), which is a fundamental parameter
in studies of WR galaxies (SV98; Leitherer 1996).
For example Contini et al. (1995) report the observations of a new WR galaxy
Mrk 712 which may require a flat IMF (Contini, Davoust, \& Consid\`{e}re
1995). Extending the sample of WR galaxies and studying both their WN and WC
content is therefore very useful to constrain the evolution of massive
stars and study the process of star formation in starbursts.



We report the discovery of two new WR galaxies from our systematic
search for WR galaxies. One of them, Mrk~1039, was 
reported with a classification as H~{\sc ii} galaxy, but the presence of 
broad emission lines in their spectra was not noticed (de Grijp et al. 
1992).
WR signatures were also found in Mrk~35 (Haro~3), previously studied by
Steel et al.
(1996). However, we observed this source in a different slit position, obtaining
different WR features from those of Steel et al.
In this paper, we present the 
observations and data reduction in \S 2, and discuss line measurements 
and the WR populations in these sources in \S 3. The implications from the 
comparisons between the observations and model predictions are discussed in 
\S 4.

\section{OBSERVATIONS}

Spectra of Mrk~35, Mrk~1039, and F08208$+$2816 were obtained under photometric 
conditions on 1997 January 10-11, February 6, November 21-23, and on 1998 
February 28 at Beijing 
Astronomical Observatory (BAO) with a 2.16m telescope. 
A grating  with 300 lines $\rm{mm}^{-1}$ yields 
a spectral resolution of 11\AA~(FWHM)
over the range of 3700-8000\AA. 
The total exposure time is 1200 sec (600$+$600 sec) for Mrk~35,
12600 sec (3$\times$3600+1800 sec) for Mrk~1039, and 10800 sec (3$\times$3600
 sec) for F08208$+$2816, respectively.
The seeing was quite stable
during the observations, at about 1.\arcsec 5 - 2.\arcsec 0.
A slit width of 2.\arcsec 5 was taken for all three sources. Mrk~35 was 
observed with a position angle of 150$^{o}$, different than the slit position
used by Steel et al. (1996), whose observations we only became aware of
recently.


The data were processed with standard IRAF procedures.
All observed line fluxes have been corrected for reddening due to the dust
in our Galaxy and within the sources themselves.
We have followed the procedures to perform the absolute flux calibrations
described in detail by Vacca \& Conti (1992).
The underlying Balmer absorption
of EW $\sim 2$\AA, which is the low-end of
the predicted value for instantaneous 
burst at a metallicity Z $<$ Z$_{\odot}$ (Olofsson 1995),
was subtracted before correcting for intrinsic extinction.
F08208$+$2816 is a unique source in this reduction, as discussed below.

The final de-reddened, rest-frame spectra are shown in Fig.~1a-1c, for Mrk~35,
Mrk~1039, and F08208$+$2816, respectively. The enlarged spectra around 4650\AA~
are also shown in the insets, in order to display the WR bump clearly. 
The broad \CIV\ lines and the broad \HeII~ lines from WR stars are
obviously present. The de-reddened line fluxes,
relative to \Hb, are given in Table 1. 
For clarity, the WR line fluxes are listed separately in Table 2.
Other measured properties are summarized in Table 3.
The metallicities have been derived using McCall's
algorithm with minor modifications (Pagel et al. 1992). McCall's
algorithm is
based on three-level-atom solutions, and hence is adequate at low and 
intermediate 
densities, with electron temperature and line ratio of 
[O {\sc iii}]($\lambda 4959+5007$)/$\lambda 4363$~
as inputs.

Mrk~1039 has also been observed with medium-resolution, centered about the
redshifted 4686\AA, with 2.16m telescope at BAO on Oct 7-9, 1997
using 600 lines $\rm{mm}^{-1}$ grating, yielding
resolution of 6.2\AA~ as measured from the FWHM of the night sky lines. The
total exposure time is 4$\times$3600 sec.
The spectra were dereddened using the internal reddening derived from 
low-resolution data listed in Table 3. The final spectrum is illustrated in 
Fig. 3.
The WR line fluxes and equivalent widths derived from medium resolution
data are also listed in Table 2, in parenthesis.

\section{WOLF-RAYET STARS IN Mrk~35, Mrk~1039, and F08208$+$2816}
 {\it Wolf-Rayet populations:}
According to the classification scheme of Smith, Shara, \& Moffat (1990$b$),
a typical value of FWHM(\CIV) for early-type WC stars is 60$^{+20}_{-10}$\AA.
The observed FWHM(\CIV) $\sim$ 70\AA~ in Mrk~35; $\sim$ 75\AA~
in Mrk~1039; and $\sim$ 71\AA~ in F08208$+$2816, are typical for WC4 stars. 
A substantial number of WO stars, i.e. comparable to the population of
WCE stars, can be discarded in these objects, considering one critical 
criterion for
the presence of WO stars is FWHM(\CIV) $>$ 90\AA, (Smith et al. 
1990$a$, Smith \& Maeder 1991). Besides, we did not detect the
O {\sc v} $\lambda  5590$~ line, 
a characteristic feature for WO1 \& 2 stars
(Barlow \& Hummer 1982). This implies that, if WO stars exist in small 
quantities, they must be WO4 stars. There is no evidence for the
presence of C {\sc iii} $\lambda 5696$~ emission line in the spectra. We can
therefore confidently exclude a major population of late type WC stars in 
Mrk~35, Mrk~1039, and F08208$+$2816.

Adopting the ionization subclass scheme for classifying WN subtypes 
(Smith et al. 1996), the early and late WN subtypes were demarcated by 
whether I(\NV) is smaller (MN6 subtype) or larger than 0.5 $\times$ 
I(\NIII) in peak fluxes. 
With this criterion, we found that the peak fluxes of
\NV\ in these three sources are below 0.5 $\times$ I(\NIII), fitting well 
to late
type WN stars. Thus, the WR populations in Mrk~35,
Mrk~1039, and F08208$+$2816 are dominated by WNL and WCE stars.

{\it Line measurements:}
The line measurement at the WR bump around 4650\AA~ is a crucial step in the 
study of WR galaxies. A problem in line measurements is severe blending
in this wavelength region. The WR bump around 4650\AA~
may in general contain broad stellar emission lines, such 
as \HeII, C {\sc iv} $\lambda 4658$, 
C {\sc iii} $\lambda 4650$, N {\sc iii} $\lambda\lambda 4634-4641$, 
N {\sc v} $\lambda 4619$, and \NV,
but also narrow nebular emission lines including \HeII, 
[Fe {\sc iii}] $\lambda 4658$, C {\sc iii} $\lambda  4650$,
N {\sc iii} $\lambda\lambda  4634-4641$, and [Ar {\sc iV}] $\lambda 4711$.
Our measurements were done using the "specfit" task
in the STSDAS software by decomposing the WR bump into the six components
(three  nebular plus three stellar) between 4640 and 4711 \AA~ listed in 
Table 2. Given the low spectral resolution and the moderate S/N of 15-20,
the uncertainties may be larger than the formal values. This is 
illustrated by differences between measurements from low-resolution
and medium-reslution observations for Mrk~1039, as we will discuss in \S 4.


\section{DISCUSSION}
{\it WR/O number ratio:}
Vacca \& Conti (1992) have developed a method to derive the WR/O number ratio 
based on the flux of \HeII~ line relative to \Hb. However, as discussed 
first by Conti (1993) and then by 
Schaerer (1996) and Schaerer \& Vacca (1998) in detail, some assumptions made
by Vacca \& Conti (1992) in deriving WR/O ratios 
render this procedure uncertain.
Among them, the evolutionary effects of Q$_{\rm o}^{\rm WR}$, the
average Lyman continuum luminosity per WR star, and the dominant
factor $\eta_{\rm o}$, a conversion factor depending on age and other
parameters, complicate the derivation of WR/O
number ratios. In SV98, these two problems have been
thoroughly investigated, supplying the evolution of Q$_{\rm o}^{\rm WR}$(t,Z) and
$\eta_{\rm o}$(t,Z) in replacing averaged Q$_{\rm o}^{\rm WR}$ and 
$\eta_{\rm o}$. 
Furthermore, uncertainties in the mean \HeII~ line luminosity
per WNL star are large, requiring further tests to verify its validity in
deriving the number of WNL stars. Despite these errors, the estimate
for the number of WNL stars using this mean luminosity is in good agreement
with the number obtained by directly counting stars (c.f. SV98).
Therefore the number
ratios of WR/O, or WC/(WR+O), WNL/(WR+O) still help us understand the
starburst phenomena. Adopting the mean \HeII/\CIV\ line luminosity of
1.6$\times$10$^{36} {\rm erg\ s^{-1}}$/3.0$\times$10$^{36} {\rm erg\ s^{-1}}$ 
and appropriate
values of Q$_{\rm o}^{\rm WR}$(t,Z) and $\eta_{\rm o}$(t,Z) from the
new model calculations of SV98, we have obtained the
number of Wolf-Rayet and O stars, and the WR/(WR+O) number ratio thereafter
in Mrk~35, Mrk~1039, and F08208$+$2816, given in Table 4.

From the comparisons of the relative number of WR stars with the model
predictions (Cervi\~{n}o \& Mas-Hesse 1994; Maeder \& Meynet 1994; 
Meynet 1995), we could see that the observed
number ratio in these three sources are well above the predicted WR/(WR$+$O)
or WR/O ratios for models with constant star formation regimes
at metallicities of (1/4 - 1/5)Z$_{\odot}$, which would be less than
0.01, or 0.03 with high mass loss. Also the observed
equivalent widths of the WR bump (see Table 2) are substantially larger
than model predictions for constant star formation 
at a metallicity of 
$\sim$ 1/5 Z$_{\odot}$, EW(WRbump) $\sim$ 0.5\AA~(Cervi\~{n}o \& Mas-Hesse 
1994).
Therefore, the event of star formation in Mrk~35, Mrk~1039, and F08208$+$2816 
would be in short burst duration. The SV98 models for constant star formation
(Version 2.32) give quite larger values for EW(WRbump) than
those in Cervi\~{n}o \& Mas-Hesse (1994). The above statement does, 
however, still hold.

In the instantaneous models, the burst age can be well indicated by the
equivalent widths of \Hb~(Copetti et al. 1986;
Maeder \& Conti 1994; Cervi\~{n}o \& Mas-Hesse 1994).
The ages obtained from the latest synthesis models (SV98)
are about 4.0Myr (Mrk~1039), 4.2Myr (Mrk~35), and 4.7Myr (F08208$+$2816).

{\it Comparisons with new synthesis models:}
SV98 have constructed new synthesis models using
up-to-date stellar evolution tracks and stellar atmospheres. The outputs 
of these models provide
us with more detailed and self-consistent predictions for WR features. 
The comparisons of our observations with the SV98 model
predictions are shown in Fig. 2, 4 and 5
for EW(\HeII)/EW(WRbump)/EW(\CIV) and their relative intensities, respectively.
The Figures show the relevant predictions for Z=0.004 and an IMF slope 
$\alpha$=1. (thick dashed line), $\alpha$=2.35 (Salpeter IMF, thick
dot-dashed), and for Z=0.001 (Salpeter IMF, thin dotted).
All these lines are drawn with the model
calculations Version 2.32 provided by Schaerer. Thick open symbols in these 
figures
denote various observations: a square for Mrk~35, a triangle
for Mrk~1039, and a circle for F08208$+$2816.
Bars over these symbols show their uncertainties listed in Table 2,
calculated following Kobulnicky \& Skillman (1996). 



Fig. 2 shows that the observations of \HeII\ (line intensities and
equivalent widths) in all objects are systematically larger than the
model predictions with a Salpeter IMF, implying that a top-heavy IMF
($\alpha$=1.0) might exist in these WR galaxies to favor the formation
of WR stars.
The large uncertainties of about 40-45\% in the
measurements, however,
make this conclusion vague (Schaerer 1997, private communication).
Particularly, as most of our data are of low resolution, the strong
broad emission features may completely dominate the wavelength region around 
4686\AA. The measured line fluxes of nebular \HeII~ are
probably underestimated, and the intensities of stellar components are most
likely overestimated.


The influence of spectral resolution on the measured nebular and stellar
lines at the 4650\AA~ region can be demonstrated by comparisons of observed
WR quantities derived from medium-resolution data, shown in Fig. 3 and listed
in parentheses in Table 2, along with the values derived from low-resolution 
data of Mrk~1039.
In fact, the line intensity of [Fe {\sc iii}] $\lambda 4658$ relative to 
\Hb=100, obtained from medium-resolution data,
is 0.9$\pm$0.4, twice as large as that estimated with low-resolution  data.
The relative line intensity of C {\sc iv} $\lambda 4658$ is reduced to
0.96$\pm$0.4 accordingly, compared with 1.1$\pm$0.6 derived from low
resolution data. The nebular and stellar \HeII~ line intensities are 
0.6/5.6 estimated from medium-resolution data, compared to 0.5/6.2 given in
Table 2. Interestingly only [FeIII] is significantly changed between the
low and medium resolution spectra. This situation, however, needs to be 
confirmed with higher resolution data.

With a spectral resolution higher than the one we used in 1997 October, 
Izotov et al. (1994, 1996,
1997$b$) found that the relative intensity of nebular \HeII~ is in the range of
$\sim$1 to 3\%. Our estimated intensity of nebular
\HeII~ is about 1\%. It is possible that a larger value of I(nebular \HeII)
will be found in these sources with data of higher resolution, and 
the relevant stellar emission will not be 
significantly larger than the model values.
Since the \HeII~ line is a dominate component in WRbump, 
the same arguments are applicable for comparisons between observed and
model EW(WRbump), shown in Fig. 4.

C {\sc iv} $\lambda 5808$ is a weak WR feature in WR galaxies, however, it is
less affected by blending than \HeII~ line in WR bump. One could
expect that
comparisons between observed and model C {\sc iv} $\lambda 5808$ lines would
provide clearer trends. Fig 5$a$ illustrates comparisons for relative line
intensities, and Fig. 5$b$ for EW(C {\sc iv} $\lambda 5808$). It is 
interesting to note that the observed EW(C {\sc iv} $\lambda 5808$) of
Mrk~1039 can be matched {\it within uncertainty} by model predictions based
on the Salpeter IMF, 
which is contradictory to the comparisons given in Fig. 2$b$ for
this source. In the mean time, the relative line intensity, 
I(C {\sc iv} $\lambda 5808$)/I(\Hb), of Mrk~1039 can be fitted by model
calculations, {\it within uncertainty}, based on flat IMF.
It should be pointed out that, 
at the burst age of 4-5 Myr for our three WR galaxies,
the continuum emission from the gas is negligible
(Leitherer \& Heckman 1995). The observed equivalent widths of WR features 
are thus insensitive to the gaseous radiation. It may, however, be reduced
by underlying stellar population. On the other hand, the relative WR line
intensities are less affected by such dilution effects, though
unrealistic comparisons for line intensities could be caused by 
the mismatch of slit width with the nebular emission regions 
(see, e.g. Meynet 1995; SV98).
Due to this possible ``light loss'' of nebular emission outside the slit,
the relative intensities of C {\sc iv} $\lambda 5808$ could be increased
by 
such a factor.

To the first
order, we might estimate the ``loss'' fraction of the Balmer emissions by an
approach suggested by 
Mas-Hesse (1997, private communication). With
a \Hb~ profile along the slit, we can estimate the fraction of \Hb~ emission
that we have accumulated for study. After taking  this ``loss'' effect of
\Hb~ emission into consideration, the observed line intensities and the 
EW(C {\sc iv} $\lambda 5808$) for the three WR galaxies are indicated by the
thin, open symbols in Fig. 5. The same corrections have been made for observed
values of \HeII~ line and the WR bump, shown also by thin, open symbols in
Fig. 2 and 4. The results obtained this way might indicate that the fitness
to observed C {\sc iv} $\lambda 5808$ in Mrk~1039 seems to move away from
model predictions based on the Salpeter IMF, as compared with those
before making correction on ``loss'' effect. For Mrk~35, however, the match
is always held, before and after this correction, to the canonical models.
F08208$+$2816 may be a source showing a clear trend of disparity
with canonical models for C {\sc iv} $\lambda 5808$.
The difference between the observed and the model predictions for this
source, however, may not be as large as indicated in the figures. 
Because of the different approaches taken,
as discussed in SV98, the duration of WC phase between the predictions of
SV98 and Meynet (1995) are different, although by use of the same stellar
models. WC phase in Meynet (1995) is longer than that in SV98.  More realistic
predictions on WC phase may be longer than that values we used for comparisons.

These comparisons with our low-resolution data tentatively suggest
a deviation from the Salpeter IMF (1955) in some
WR galaxies, which are worth studying at higher resolution. We are
planning to do so, especially considering the fact that
both \HeII~ and C {\sc iv} $\lambda 5808$ are detected in these WR galaxies.

The number ratio of WC/WN derived from these two features can provide critical
tests on the mass-loss rates, 
a key input for stellar evolution models of massive stars
(Maeder \& Meynet 1994, Meynet 1995), which has also important
consequences for population synthesis models.
Indeed, different mass-loss rates can produce different distributions of WR 
stars among WC and WN subtypes. For example, the number ratio of WC/WNL $<$ 1
has been predicted by models of instantaneous starburst with standard mass
loss rates at burst age of 4 Myr for a metallicity of Z=0.004, while WC/WNL
$>$ 1 by models with high mass-loss rates (Meynet 1995). SV98 models are the 
ones with high mass-loss rates.
The observed number ratios
WC/WNL for Mrk~35/Mrk~1039/F08208$+$2816 are 0.33/0.34/0.42 at age of
3.9/3.6/4.4 Myr 
(newly derived ages corrected for the `loss' of nebular emission as discussed
above, also see Table 4), respectively. 
To be precise, the number of WNL stars should
be derived from the luminosity of observed \HeII~ after subtraction of the
contribution from WC stars to \HeII\ (de Mello et al. 1998).
In the compilation of WR line luminosities in SV98, this contribution is 
included in 4650 blend. For WC4 stars, the \HeII~ from WC stars contributes
8-30\% of the 4650 blend, based on the estimation of Smith et al. (1990a).
The 'precise' number of WNL stars could be estimated from the luminosity of
(L$_{\rm obs}$(\HeII)-N(WC)$\times frac \times 1.71\times 3.0\times 10^{36}$),
where $frac$ denotes the fraction of 4650 blend being the contribution of 
\HeII~ from per WC star,
and 1.71 is the mean flux ratio of 4650 blend with respect to 
C {\sc iv} $\lambda 5808$ for WC4 star, adopting from Table 2 of SV98.
The values listed in parentheses in the row 'WNL' of Table 4 are those 
'precise' number of WNL stars, taking $frac$=19\%.
The newly derived number ratios then become
0.42/0.43/0.57, respectively, suggesting no need of high mass-loss rates
for these three WR galaxies. 
The highest number ratio of WC/WNL, in the case of F08208$+$2816 and
$frac$=30\%, is 0.72.
To change high mass-loss rates to standard ones will significantly reduce
the formation of WR stars and the duration of WR phase, as discussed by
Meynet (1995), deteriorating the comparisons with our observations shown in 
Fig 2, 4 and 5.
Due to a probable overestimation of line
intensities in \HeII~  discussed above, more realistic WC/WNL 
number ratios
will be derived from data of higher resolution, and
the unsettled dilemma might be eased.
We may find new
constraints on models suitable for the new WR galaxies reported here.


It is worth mentioning the detection of He {\sc ii} $\lambda 5411$
(FWHM $\sim$ 24\AA) with 3 $\sigma$
in Mrk~1039, the first detection of this broad line from WR stars 
outside the Local Group. 
The detection of this line might be related to 
the exceptionally strong \HeII~ in Mrk~1039. Compared to the  
EW(\HeII) in Mrk~1039, those in Mrk~35 and F08208$+$2816 are about half as much
(Fig. 2), and we have found no evidence of 
He {\sc ii} $\lambda 5411$~ in these two sources. 


{\it F08208$+$2816:}
F08208$+$2816 is a luminous IRAS galaxy. Its infrared
luminosity is L(8-1000$\mu m$)=1.4$\times$10$^{11}$L$_{\odot}$, following 
Sanders \& Mirabel (1996). If we take the Balmer decrements 
as a reddening tracer, 
then F08208$+$2816 has little internal reddening by
dust. Indeed, the observed F(\Ha)/F(\Hb) ratio fits well, within uncertainty,
to the theoretical I(\Ha)/I(\Hb) derived from Case B recombination. Also, 
the correction for the underlying Balmer absorption for this source
would not need to be greater than 0.25\AA.

Fig. 6 shows the field centred about F08208$+$2816, extracted from the
Digitized Sky Survey. 
The WR features are detected in Source 2. Source 4 is a foreground star.
Source 3 is too faint for the 2.16m telescope at BAO. Source 1 is faint, and
shows properties of a H {\sc ii} galaxy with the same redshift as Source 2.
Apparently, F08208$+$2816 is a merging system.
It is therefore another example of a luminous IRAS galaxy (LIRG)
indicating the merging process as a triggering
mechanism for starburst (Sanders et al. 1988). And it is the second LIRG
with detection of
WR stellar population or young massive O stars,
providing direct evidence for the connection between starburst and very
powerful far-infrared emission. The other two sources are IRAS 01003$-$2238
(a WR galaxy, Armus, Heckman \& Miley 1988), and IRAS 08339$+$6517 
(Gonzalez-Delgado et al. 1998). 
The luminosity of L(WRbump)(F08208$+$2816)=3.8$\times 10^{40}$ ergs s$^{-1}$,
about a decade lower than that in IRAS 01003$-$2238 (Armus et al. 1988).
Interestingly, F08208$+$2816 is the only source in these three LIRG
that shows essentially zero internal reddening, as derived from the Balmer
decrements.

It is possible that some of the dust clouds responsible for infrared emission
could be located close to the \Ha-emitting gas, as suggested by 
Ma\'{\i}z-Apell\'{a}niz et al. (1998) for their observations of NGC4214. The
scattering of \Ha~ photons would reduce the Balmer ratio. Alternatively, 
the stellar winds from massive stars could expel the interstellar medium
and the region we observed might then be density bounded instead of ionization
bounded. 

In view of the existence of a large number of young massive stars in Source 2
of F08208$+$2816 ($ > 2 \times 10^4$ WR stars, or $\sim 1.4 \times 10^5$ 
massive O stars) a
powerful superwind could be expected in F08208$+$2816. Therefore the 
interstellar
medium in Source 2 has probably been blown off. The region we studied is likely
not ionization bounded, implying that a large fraction of the Lyman-continuum
photons of 1.4$\times$10$^{54}$/sec might escape from Source 2 of 
F08208$+$2816. Indeed, IRAS 08339$+$6517 is a source with about the same
infrared luminosity as F08208$+$2816, its internal reddening derived from
the Balmer decrements is 0.55 mag (Leitherer et al. 1995). The escape fraction 
of Lyman continuum photons, estimated from the observed flux at about 900\AA~
with the Hopkins Ultraviolet Telescope (HUT), is $<$ 4.1\% (Hurwitz et al. 
1997).
Another starburst galaxy observed with HUT, Mrk~66, shows same zero internal
reddening as Source 2 of F08208$+$2816, but with lower infrared
luminosity, about five times less than that of F08208$+$2816. It is 
interesting to see that the upper limit of escape fraction of Lyman photons
from Mrk~66 is at least 43\%  (cf. Hurwitz et al. 1997). 
Based on these new results, F08208$+$2816 might be
a promising source for understanding the contribution from
young starbursts to the UV background radiation, especially considering
that merging systems should have been more frequent in the early universe.


\acknowledgements{We would like to thank an anonymous referee for his
thoughtful and instructive comments on our paper. A substantial revision
was made following his very detailed suggestions.
We are grateful to D. Kunth, M. Mas-Hesse and F. Legrand
for helpful discussions and suggestions
on an earlier version of the manuscript. We also thank D. Schaerer for
providing us with SV98 models prior to publication, and for his critical
comments on our later version, which strongly deepened our analyses.
The BAO staffs are thanked for their assistance during the observations.
This work is supported by a grant from the NSF of China,
and a grant from the Ascent Project of the State Science Commission of China.}

\clearpage
\begin{deluxetable}{clrrr}
\footnotesize
\tablecaption{Properties of the Spectral Lines}
\tablewidth{0 pt}
\tablehead{
\colhead{$\lambda_{0}$} &\colhead{Ion}  &
\multicolumn{3}{c}{Intensity\tablenotemark{a}} \\
\colhead{(\AA)}&&\colhead{Mrk~35}&\colhead{Mrk~1039}&\colhead{F08208$+$2816}
}
\startdata
3727    &  [O {\sc ii}]&155.0  & 98.0&151.2 \nl
3869 &  [Ne {\sc iii}]+He {\sc i} &22.8&20.3&29.6\nl
3967 & \HE+[Ne {\sc iii}] &26.9&18.8&18.5\nl
4101& \Hd &25.1& 14.7&20.9\nl
4340& \Hr &46.0& 35.3&42.8\nl
4363 & [O {\sc iii}]&2.3 &4.3&3.5\nl
4471 & HeI &4.3 &4.3&3.5\nl
4861 &\Hb & 100.0&100.0&100.0\nl
4959&[O {\sc iii}]&130.1 &158.2&134.1\nl
5007&[O {\sc iii}]&396.0&475.9&404.5\nl
5200 & [N {\sc ii}]&0.6 & 1.8&2.3\nl
5876 & He {\sc i} &13.1&15.1 &13.2\nl
6304&[O {\sc i}]+[S {\sc iii}]&4.1&6.0 &7.1\nl
6563&\Ha&286.6& 282.9&282.0\nl
6584&[N {\sc ii}]&16.9& 14.5&33.6\nl
6716&[S {\sc ii}]&13.9&16.2&24.7\nl
6731&[S {\sc ii}]&13.9&13.4&19.2
\enddata
\tablenotetext{a}{De-reddened line fluxes relative to I(\Hb) = 100}
\end{deluxetable}

\newpage
\begin{deluxetable}{clrrr}
\footnotesize
\tablecaption{Wolf-Rayet Features}
\tablewidth{0 pt}
\tablehead{
\colhead{$\lambda_{0}$} &\colhead{Ion}  &
\multicolumn{3}{c}{Intensity\tablenotemark{a}} \\
\colhead{(\AA)}&&\colhead{Mrk~35}&\colhead{Mrk~1039}&\colhead{F08208$+$2816}
}
\startdata
4640&N {\sc iii}&1.0$\pm$0.6&0.7$\pm$0.4&1.1$\pm$0.7\nl
    &&&(0.5$\pm$0.3)\tablenotemark{b}&\nl
4658\tablenotemark{n}&[Fe {\sc iii}]&0.2$\pm$0.1&0.4$\pm$0.2&0.8$\pm$0.5\nl
&&&(0.9$\pm$0.4)&\nl
4658\tablenotemark{s}&C {\sc iv}&1.8$\pm$1.0&1.1$\pm$0.6&1.2$\pm$0.7\nl
&&&(0.96$\pm$0.4)&\nl
4686\tablenotemark{n}&He {\sc ii} &0.2$\pm$0.1 &0.5$\pm$0.3&...\nl
    &&&(0.6$\pm$0.3)\nl
4686\tablenotemark{s}&He {\sc ii} &4.0$\pm$2.2 &6.2$\pm$2.5&5.6$\pm$3.0\nl
    &&&(5.6$\pm$2.2)\nl
4711 & [Ar {\sc iv}] &0.6$\pm$0.4&0.7$\pm$0.4&1.0$\pm$0.6\nl
    &&&(0.5$\pm$0.3)&\nl
5411&He {\sc ii}&...&1.0$\pm$0.2&...\nl
5808 & C {\sc iv} &2.5$\pm$0.8 &3.9$\pm$1.2&4.5$\pm$1.7\\
&&&&\nl
\multicolumn{2}{c}{EW(He {\sc ii}4686)}&6.0$\pm$3.3&10.4$\pm$4.2&4.5$\pm$2.5\nl
\multicolumn{2}{c}{}&&(9.9$\pm$3.9)&\nl
\multicolumn{2}{c}{EW(C {\sc iv}5808)}&5.3$\pm$1.7&7.7$\pm$2.3&5.3$\pm$1.7\nl
\multicolumn{2}{c}{EW(WRbump)}&9.8$\pm$5.5&14.8$\pm$6.7&6.4$\pm$3.6\nl
\multicolumn{2}{c}{}&&(12.6$\pm$5.6)&
\enddata
\tablenotetext{a}{De-reddened line fluxes relative to I(\Hb)=100}
\tablenotetext{b}{The values listed in parenthesis are those obtained from medium resolution data}
\tablenotetext{n}{Nebular, narrow emission lines}
\tablenotetext{s}{Stellar, broad emission lines}
\end{deluxetable}

\begin{deluxetable}{lrrr}
\footnotesize
\tablecaption{Properties of \Hb\ and Reddening}
\tablewidth{0 pt}
\tablehead{
  & \colhead{Mrk~35} & \colhead{Mrk~1039}&\colhead{F08208$+$2816}
} 
\startdata
z&0.00327&0.00656&0.0472\nl
EW(\Hb) (\AA)&145.7&174.8&80.9\nl
I(\Hb) ($10^{-13}$ \ergscm)&6.17&6.36&1.14\nl
$[$O/H$]$&8.32&8.17&8.15\nl
E(B-V) & 0.112&0.469&$\sim$0 \nl
Distance (Mpc)\tablenotemark{a}& 13.1&26.2&188.8
\enddata
\tablenotetext{a}{H$_{\rm o}$ = 75~\Hubble}
\end{deluxetable}



\begin{deluxetable}{lrrr}
\footnotesize
\tablecaption{WR/(WR$+$O) number ratios}
\tablewidth{0 pt}
\tablehead{
 & \colhead{Mrk~35} & \colhead{Mrk~1039}   & \colhead{F08208+2816}   
} 
\startdata
WC&104&687&7213\nl
WNL&313&2018&1.695$\times 10^{4}$\nl
   &(250)\tablenotemark{a}&(1602)\tablenotemark{a}&(1.258$\times 10^{4}$)\tablenotemark{a}\nl
O&3620&9720&1.895$\times 10^{5}$\nl
&(2795)&(6430)&(1.419$\times 10^{5}$)\nl
WC/(WR$+$O)&0.026&0.055&0.034\nl
	&(0.033)&(0.079)&(0.045)\nl 
WNL/(WR$+$O)&0.077&0.16&0.079\nl
	&(0.079)&(0.18)&(0.078)\nl
WR/(WR$+$O)&0.10&0.22&0.11\nl 
	&(0.11)&(0.26)&(0.12)\nl
burst age (Myr)&4.2&4.0&4.7\nl
	&(3.9)\tablenotemark{b}&(3.6)\tablenotemark{b}&(4.4)\tablenotemark{b}
\enddata
\tablenotetext{a}{number of WNL stars after subtracting the contribution of
WC stars to 4686\AA~ line. The corrected new values of O stars and the number 
ratios are all listed in parentheses thereafter.}
\tablenotetext{b}{age corrected for ``light loss'' of nebular emission}
\end{deluxetable}

\clearpage

\noindent Figure captions:

\noindent Figure 1$a$. Spectrum of  Mrk~35 with low resolution,
where two WR bumps
around 4650\AA~ and 5808\AA~ are marked. The inset shows enlargement 
on the WR bumps around 4600\AA, where the broad \HeII~ line from
Wolf-Rayet stars is positively detected.

\noindent Figure 1$b$. Same as Fig 1a, but for Mrk~1039.

\noindent Figure 1$c$. Same as Fig 1a, but for F08208$+$2816.

\noindent Figure 2. Comparisons between the predicted and observed quantities
of stellar \HeII~ line
for Z=0.004 (dotted-dashed), 0.001 (thin dotted) based on Salpeter IMF, and 
Z=0.004 (thick dashed) based on a very flat IMF ($\alpha$=1.0),
a) relative line intensities; b) equivalent widths.
Observed data are
denoted by thick open symbols: square for Mrk~35; triangle for Mrk~1039;
and circle for F08208$+$2816. 
See the text for the thin open symbols.

\noindent Figure 3. Spectrum of Mrk~1039 with medium resolution 
taken in the blue
wavelength band. The inset shows enlargement on the WR bump as those in Fig 1.

\noindent Figure 4. Comparisons between the predicted and observed equivalent
widths of WR bump. Same notations used as those in Fig 1.

\noindent Figure 5. Same as Fig 1, but for C {\sc iv} $\lambda 5808$ line.

\noindent Figure 6. Field centered about F08208$+$2816, in 
4$^{\prime}\times$4$^{\prime}$. North is up, and east is to the left.
Sources 1-3 are three components in F08208$+$2816 system. Source 2 is
where WCE and WNL stars are detected. Source 4 is a foreground star. 
\clearpage

\begin{figure*}
\figurenum{1a}
\plotone{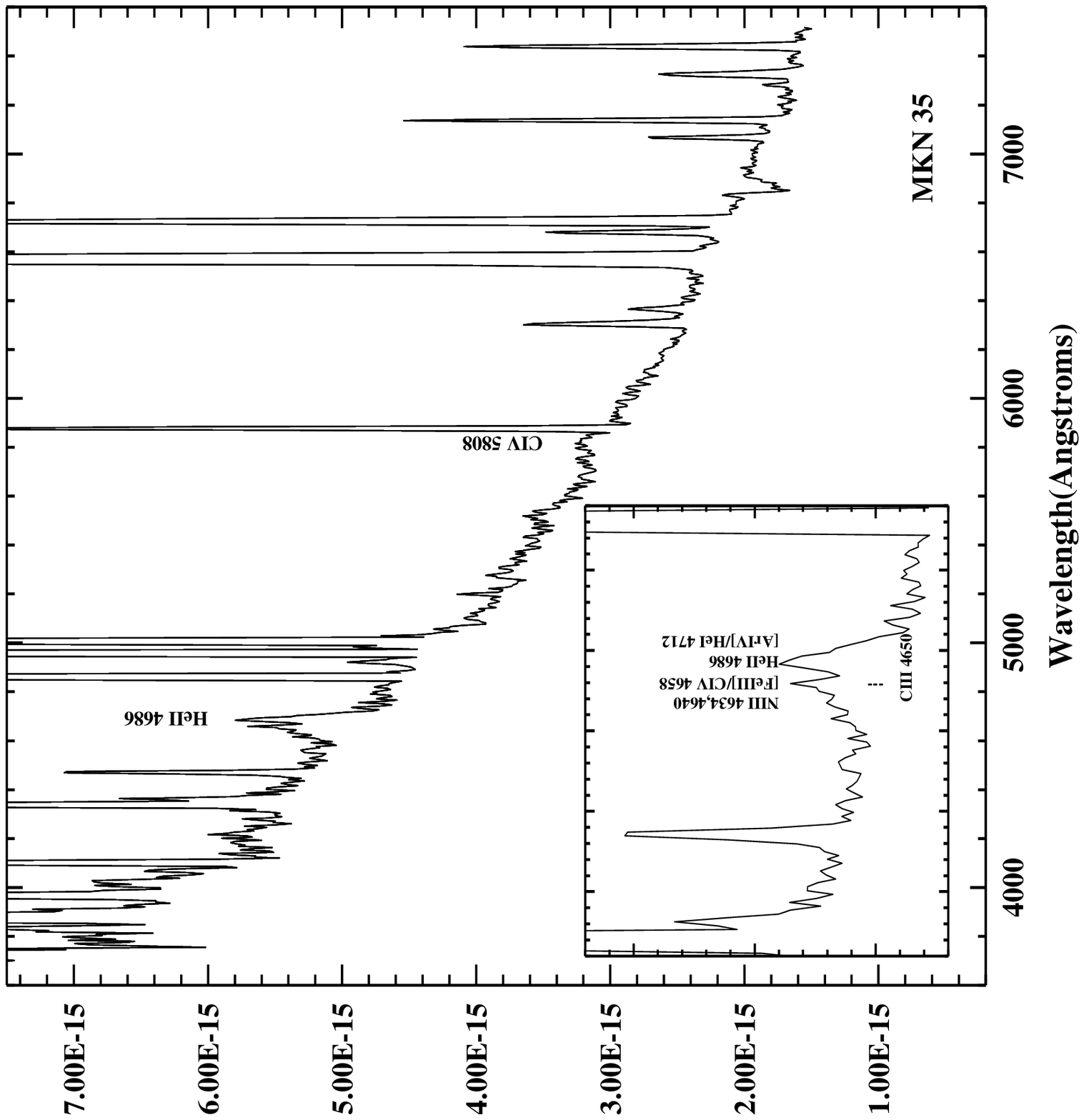}
\figcaption{
\label{Fig 1a}}
\end{figure*}

\begin{figure*}
\figurenum{1b}
\plotone{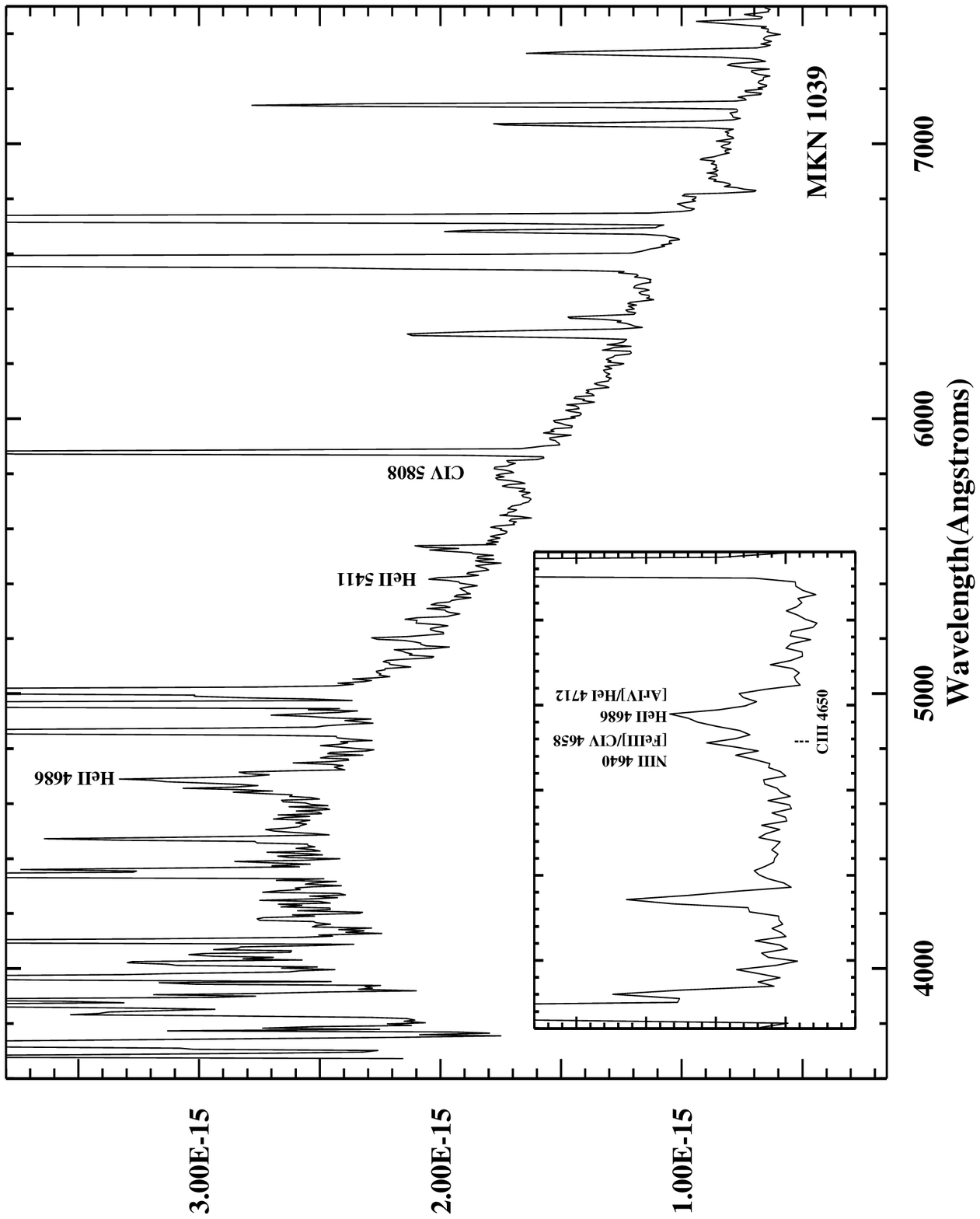}
\figcaption{
\label{Fig 1b}}
\end{figure*}

\begin{figure*}
\figurenum{1c}
\plotone{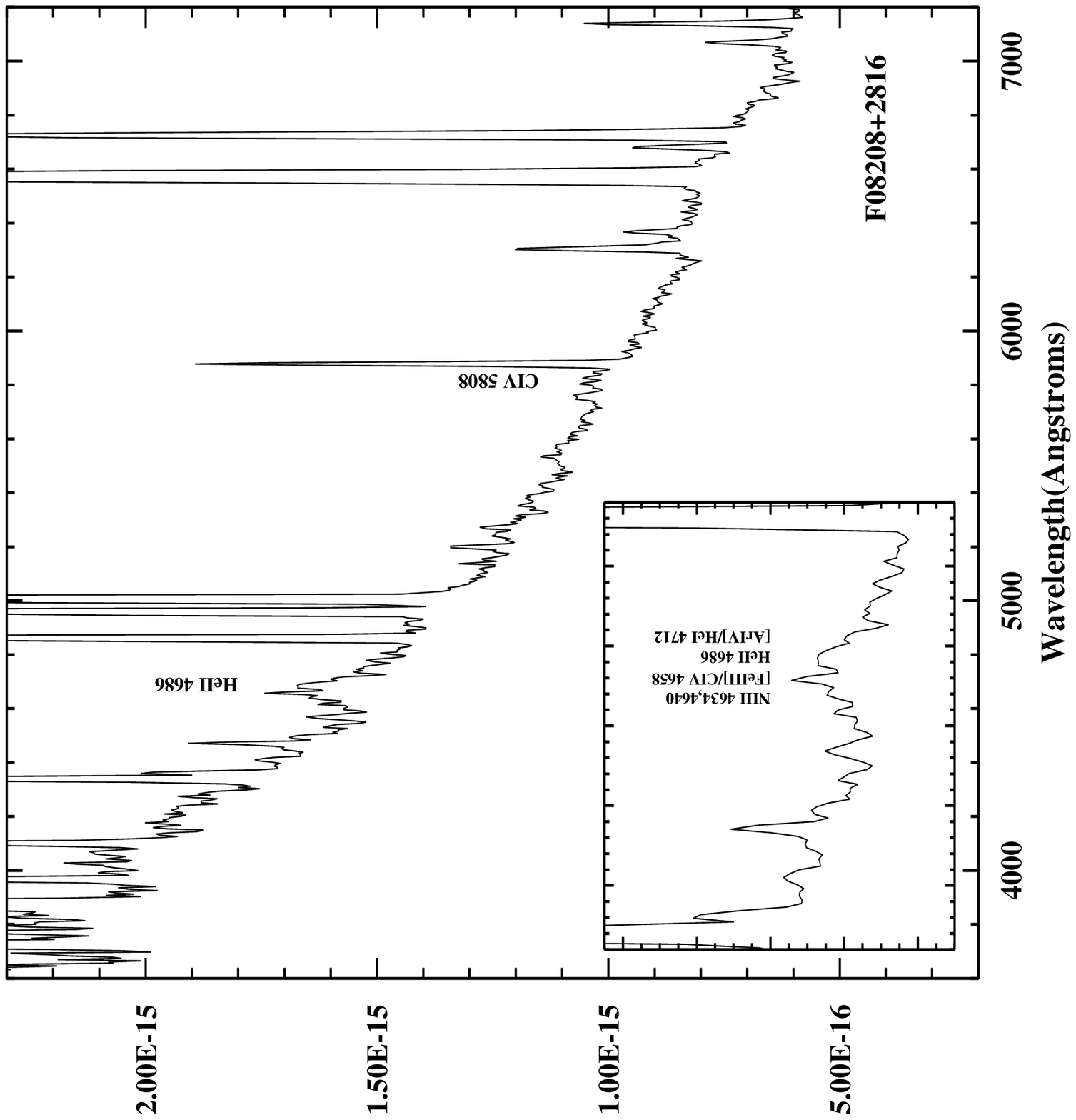}
\figcaption{
\label{Fig 1c}}
\end{figure*}

\begin{figure*}
\figurenum{2}
\plotone{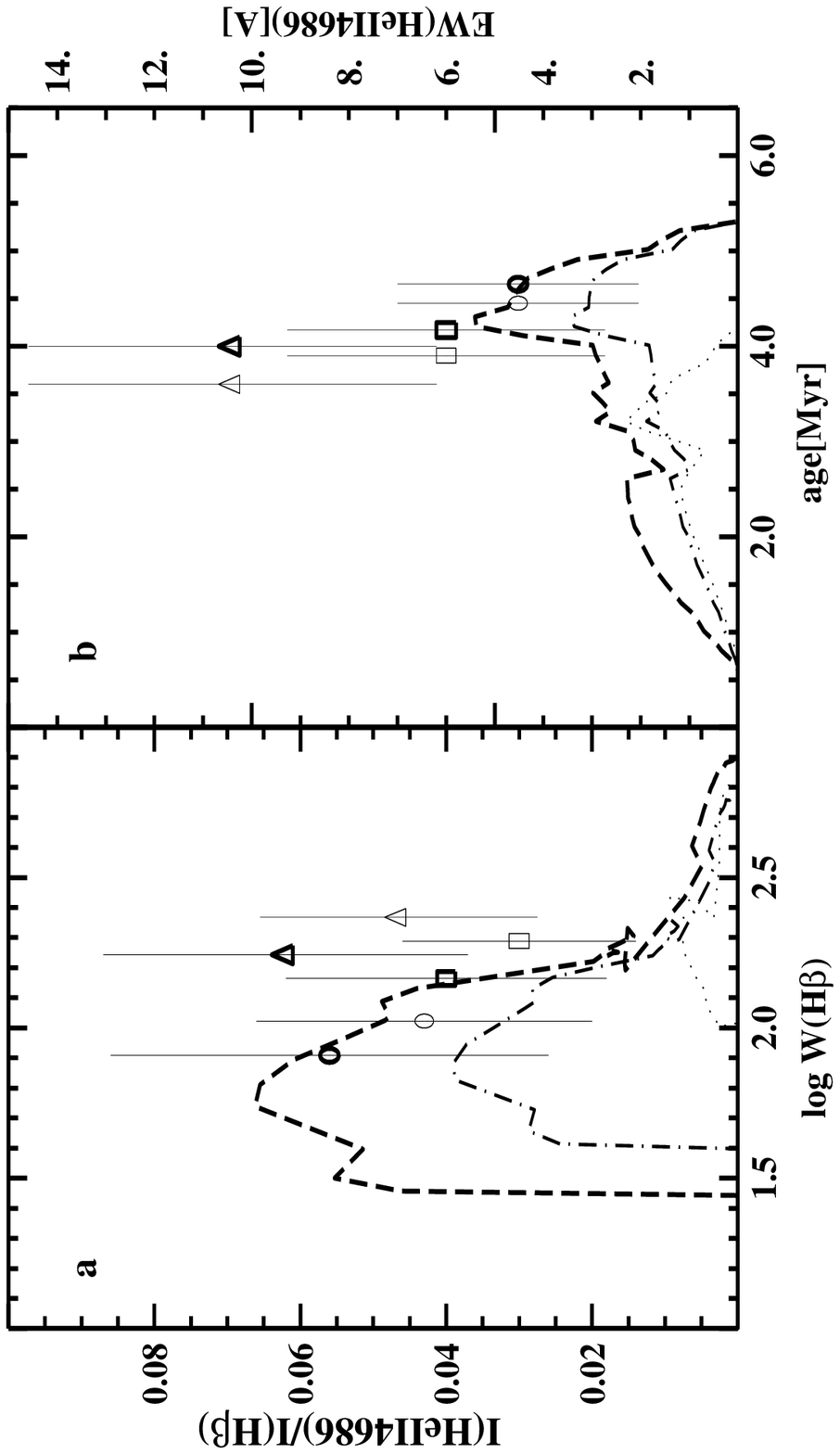}
\figcaption{
\label{Fig 2}}
\end{figure*}

\begin{figure*}
\figurenum{3}
\plotone{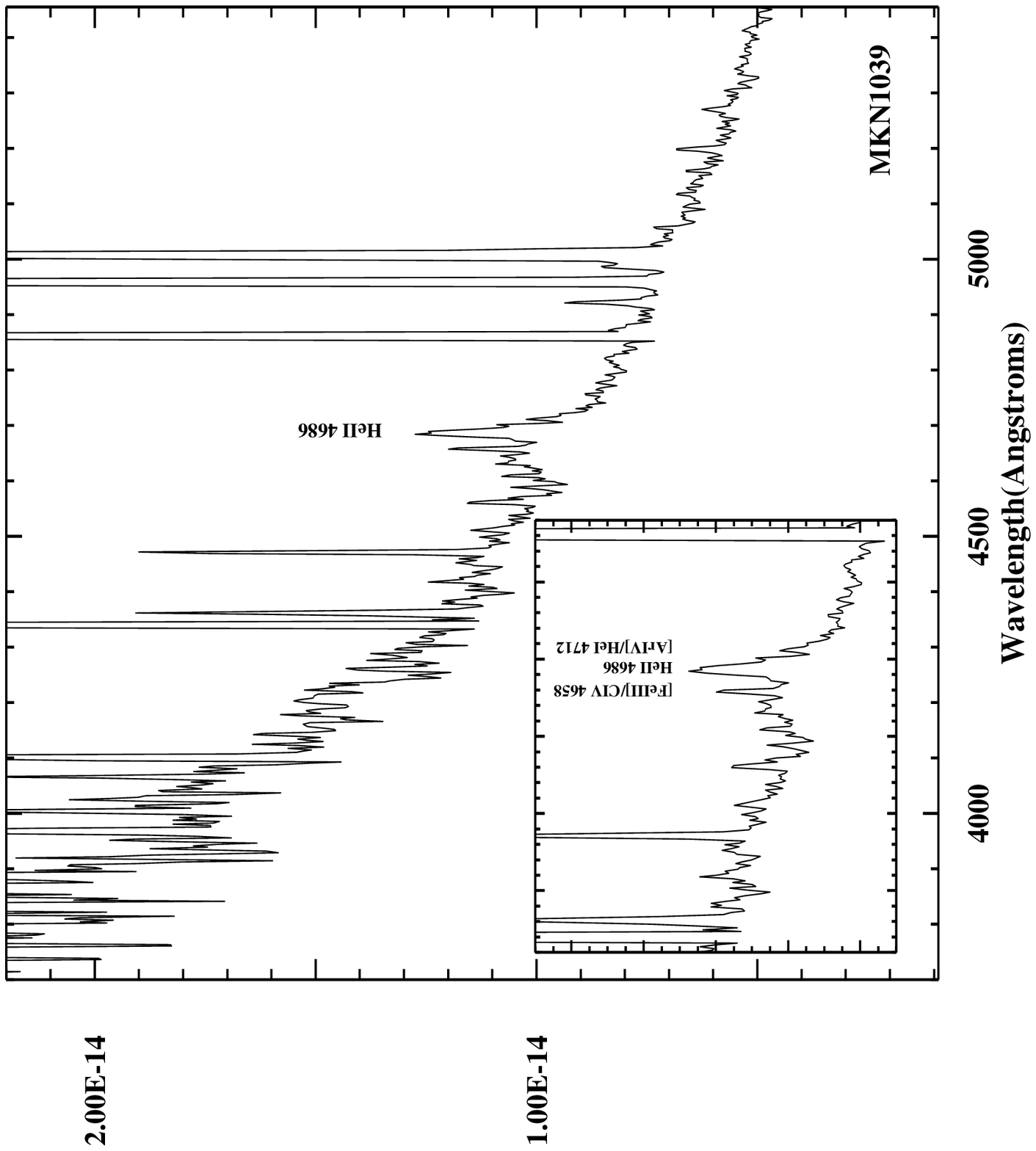}
\figcaption{
\label{Fig 3}}
\end{figure*}

\begin{figure*}
\figurenum{4}
\plotone{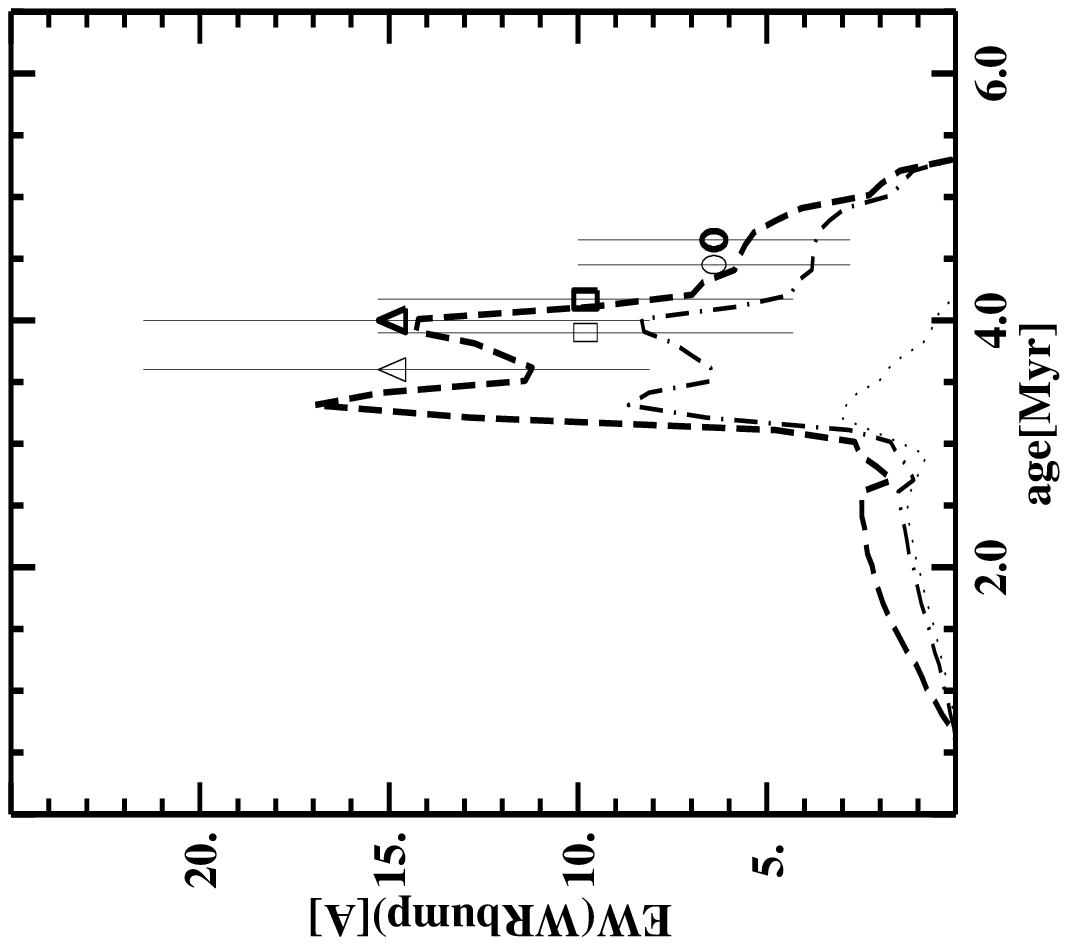}
\figcaption{
\label{Fig 4}}
\end{figure*}

\begin{figure*}
\figurenum{5}
\plotone{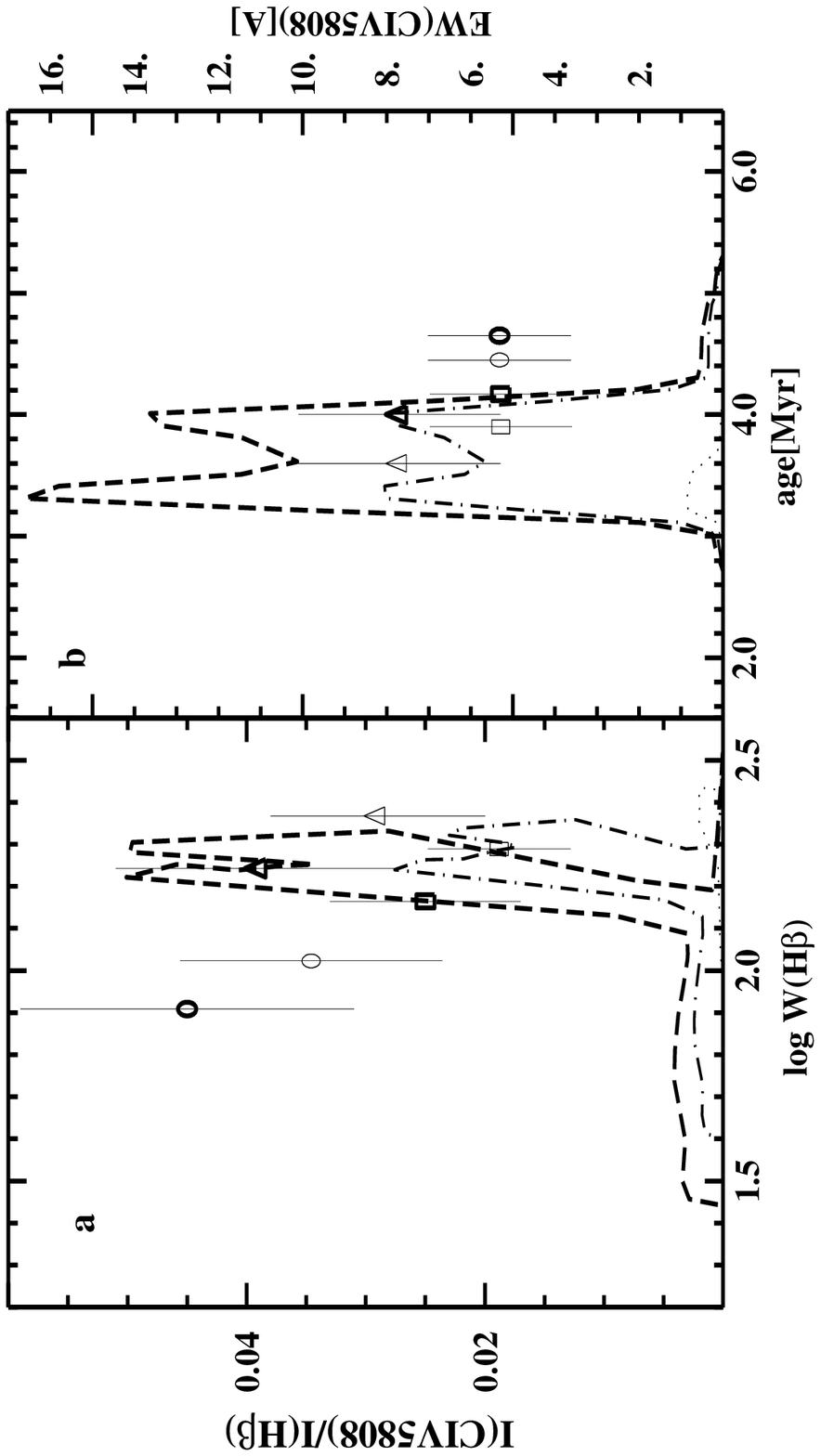}
\figcaption{
\label{Fig 5}}
\end{figure*}

\begin{figure*}
\figurenum{6}
\plotone{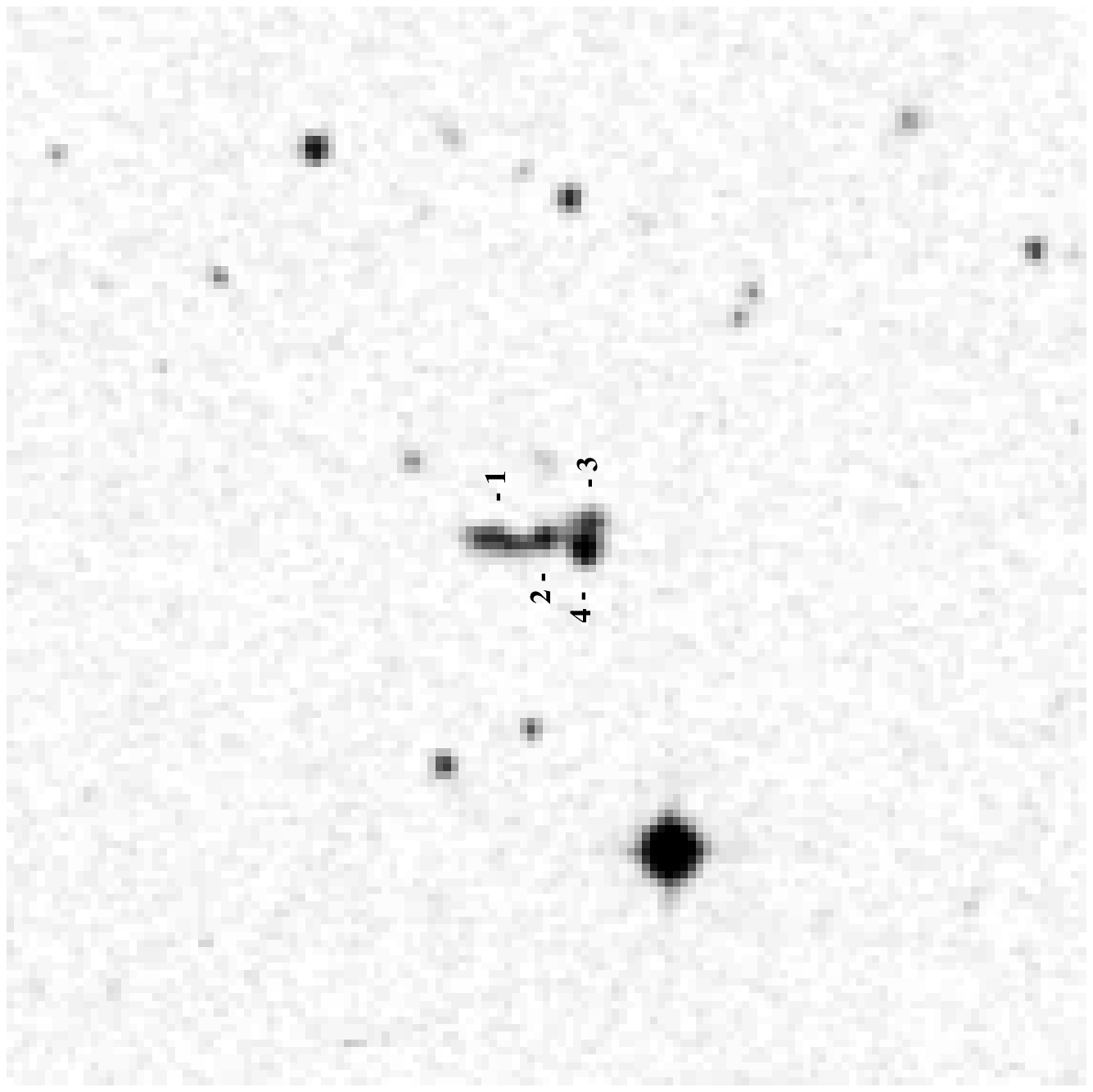}
\figcaption{
\label{Fig 6}}
\end{figure*}

\clearpage


\clearpage
\begin{thebibliography}{}
\bibitem{} Armus, L., Heckman, T.M., \& Miley, G.K. 1988, ApJ, 326, L45
\bibitem{} Barlow, M.J. \& Hummer, D.G. 1982, in Wolf-Rayet Stars:
	Observations, Physics, Evolution, ed. C. W. H. de Loore, A. J. Wills,
	(Reidel: Dordrecht), 387
\bibitem{} Cervi\~{n}o, M., \& Mas-Hesse, J.M. 1994, A\&A, 284, 749
\bibitem{} Conti, P.S. 1991, ApJ, 377, 115
\bibitem{} Conti, P.S. 1993, in Massive Stars: Their lives in the interstellar
	medium, ASP Conf. Series 35, p.449
\bibitem{} Contini, T., Davoust, E., \& Consid\`{e}re, S. 1995, A\&A, 303, 440
\bibitem{} Copetti, M.V.F., Pastoriza, M.G., \& Dottori, H.A. 1986, A\&A, 156, 111
\bibitem{} de Grijp, M.H.K., Keel W.C., Miley G.K., Goudfrooij P., \& Lub, J.
	1992, A\&AS, 96, 389
\bibitem{} de Mello, D.F., Schaerer, D., Heldmann, J., \& Leitherer, C. 1998,
	ApJ, in press.
\bibitem{} Gonz\'{a}lez Delgado, R.M., Leitherer, C., Heckman, T., Lowenthal,
	J.D., Ferguson, H.C., \& Robert, C. 1998, ApJ, 495, 698
\bibitem{} Hurwitz, M., Jelinsky, P., \& Van Dyke Dixon, W. 1997, ApJ, 481, L31
\bibitem{} Izotov, Y.I., Dyak,A.B., Chaffe, F.H., Foltz, C.B., Kniazev, A.Y.,
	\& Lipovetsky, V.A. 1996, ApJ, 458, 524
\bibitem{} Izotov, Y.I., Foltz, C.B., Green, R.F., Guseva, N.G., \& Thuan, T.X.
	1997$a$, ApJL, 487, 37 
\bibitem{} Izotov, Y.I., Thuan, T.X., \& Lipovetsky, V.A. 1994, ApJ, 435, 647
\bibitem{} \underline{\hskip 7 em}. 1997$b$, ApJS, 108, 1  
\bibitem{} Kobulnicky, H.A., \& Skillman, E.D. 1996, ApJ, 471, 211
\bibitem{} Legrand, F., Kunth, D., Roy, J.-R., Mas-Hesse, J.M., \& Walsh, J.R.
	1997, A\&A, 326, L17
\bibitem{} Leitherer, C. 1996, STScI preprint No. 1175
\bibitem{} Leitherer, C., Ferguson, H.C., Heckman, T.M., \& Lowenthal, J.D. 
	1995, ApJ, 454, L19
\bibitem{} Leitherer, C., \& Heckman, T.M. 1995, ApJS, 96, 9
\bibitem{} Maeder, A., \& Conti, P.S. 1994, ARAA, 32, 227
\bibitem{} Maeder, A., \& Meynet, G. 1994, A\&A, 287, 803
\bibitem{} Ma\'{\i}z-Apell\'{a}niz, J., Mas-Hesse, J.M., 
   Mu\~{n}oz-Tu\~{n}\"{o}n, C., V\'{\i}lchez, J.M., \& Casta\~{n}eda, H.O. 
   1998, A\&A, 329, 409 
\bibitem{} Meynet, G. 1995, A\&A, 298, 767
\bibitem{} Olofsson, K. 1995, A\&AS, 111, 57
\bibitem{} Pagel, B.E.J., Simonson, E.A., Terlevich, R.J., \& Edmunds, M.G.
	1992, MNRAS, 255, 325
\bibitem{} Salpeter, E.E. 1955, ApJ, 121, 161
\bibitem{} Sanders, D.B., \& Mirabel, I.F. 1996, ARAA, 34, 749
\bibitem{} Sanders, D.B., et al. 1988, ApJ, 325, 74
\bibitem{} Schaerer, D. 1996, ApJ, 467, L17
\bibitem{} Schaerer, D., \& Vacca, W.D. 1998, ApJ, 497, 618 
  (SV98)
\bibitem{} Schaerer, D., Contini, T., Kunth, D., \& Meynet, G. 1997, ApJ, 481, L75
\bibitem{} Smith, L.F., \& Maeder, A. 1991, A\&A, 241, 77
\bibitem{} Smith, L.F., Shara, M.M., \& Moffat, A.F.J. 1990$a$, ApJ, 348, 471
\bibitem{} \underline{\hskip 7 em}. 1990$b$, ApJ, 358, 229 
\bibitem{} \underline{\hskip 7 em}. 1996, MNRAS, 281, 163
\bibitem{} Steel, S.J., Smith, N., Metcalfe, L, Rabbette, M., \& McBreen, B.
	1996, A\&A, 311, 721
\bibitem{} Vacca, W.D., \& Conti, P.S. 1992, ApJ, 401, 543
\end{thebibliography}
\end{document}